\documentclass[twocolumn,twoside,slac_two]{revtex4}
\usepackage{graphicx}
\usepackage{fancyhdr}
\usepackage{hyperref}
\hypersetup{
    colorlinks=true,
    urlcolor=magenta,
}
\begin{document}

\title{Dark matter overview}

%

\author{N. Fornengo}
\affiliation{University of Torino, Department of Physics and INFN/Torino, via Pietro Giuria 1, 10125 Torino, Italy}

\begin{abstract}
The identification of a solution to the dark matter problem has many arrows to its bow: if dark matter is a new elementary particle, both laboratory experiments and astrophysics can bring relevant and complementary pieces of information, that than can be confronted and composed to solve this intriguing puzzle. Although we currently do not have a unique and obvious target for the DM particle, we can rely on a broad range of ideas, tools and methods that make the investigation of dark matter a multi-frequency, multi-messenger and multi-techniques integrated endeavour.
\end{abstract}

\maketitle

\thispagestyle{fancy}


The presence of dark matter (DM) is supported by copious and consistent astrophysical and cosmological observations \cite{DM}. On large scales, the average DM density is about 6 times the baryon density \cite{Planck2015}. On smaller scales, the DM distribution is quite anisotropic and hierarchical, comprising clusters, galaxies and (at least in the cold dark matter scenario \cite{NFW}) galactic substructures. Observations are consistent with a theoretical understanding of cosmic structure formation through gravitational instability, based on the LCDM model \cite{NFW}. Even though some issues are under discussion on small (galactic and sub-galactic) scales \cite{problems}, including the presence of a core in the dark matter distribution in galaxies \cite{salucci}, the basic picture appears to be consistent. The role of baryons in galaxy formation just started to be investigated recently, and it might well offer solutions to some (or all) of the small scale shortcomings \cite{solutions}.

Up to know, the evidence for DM is purely of gravitational origin: galaxy cluster dynamics, rotational curves of spiral galaxies, gravitational lensing, hydrodynamical equilibrium of hot gas in galaxy clusters, the energy budget of the Universe and the same theory of structure formation are all based on gravity and they all require the presence of DM. This implies that what we call ``dark matter'' could be either a manifestation of our incomplete understanding of gravity on scales larger than the solar system, or to a new fundamental component of Nature, namely a new elementary particle, not yet discovered and not present in the Standard Model of fundamental interactions. 
If this is the correct solution to the DM problem, it implies that the Standard Model of elementary particles needs to be extended and New Physics is required in terms of new particles. However, to demonstrate that this is indeed the correct solution to the DM problem, 
a non-gravitational signal (due to its particle physics nature) is needed.

To this aim, we can rely on multiple approaches, that well complement each other. From one side, we can produce new particles in the laboratory, including the new DM particles or associated states that are present in the underlying New Physics model. One thing that poses a great challenge is the fact that, although we have a typical target for the size of the DM particle interactions (weakly interacting with ordinary matter, or with itself, although way outs are possible -- even though the origin of the DM particle interactions is not the weak sector of the Standard Model or a direct extension of it, nevertheless the DM particle needs to be significantly weakly coupled, like e.g. in the case of millicharged DM), we have no direct indication on the mass scale of the DM particle: it can range from the $\mu$eV scale of axions (or even smaller, for suitable axion-like-particles) way up to the Planck scale, with special interesting ranges at the keV scale (like, e.g. for non-thermal sterile-neutrino warm dark matter) and the canonical GeV to few-TeV range for WIMPs. This implies that searches in the laboratory need to be performed both at the high-energy frontier, today represented by the LHC, and at the lower-energy but high-intensity frontier, like e.g. in beam dump experiments or in experiments dedicated to axions.
In both cases, the search for New Physics in the laboratory would allow to identify the physical properties of the DM sector, in a controlled environment.

The other side of the DM investigation looks instead to the DM particle in its astrophysical environment, where it actually is known to be present. In this case a whole and extensive host of astrophysical signals is under intense investigation: we can make full use of a wide and complex multi-messenger, multi-wavelength and multi-technique strategy, that points at identifying astrophysical messengers produced by the DM particle, or at revealing astrophysical manifestation its particle physics nature. Although signals are not produced under our own control (like in the case of the laboratory investigation) and complex astrophysical backgrounds are present, however in this approach DM and its particle nature are tested in  its natural environment. 

\section{Astrophysical messengers of dark matter}

Depending on its mass and interactions, DM can produce a large variety of astrophysical messengers, than can reach us and therefore be used to probe its particle physics nature. Let us recall that DM needs to be dynamically relatively cold, in order to form the observed large scale structure of the Universe: this implies that DM signals produced by DM annihilation cannot have energies larger than the DM mass (or half its mass, in case of DM decay). This fact sets the target of investigation for the different DM candidates. WIMPs, with masses in the GeV to few TeV range, can produce energetic neutrinos, photons in the gamma-ray band and energetic electrons and positrons. They can also produce an exotic component of antiprotons and heavier antinuclei, most notably antideuterium \cite{antiD} and antihelium \cite{antiHe}. Moreover, whenever WIMPs produce electrons and positrons, these latter can suffer secondary electromagnetic effects, and produce an additional component of gamma-rays, as a consequence of the up-scatter of CMB and 
other radiation fields through inverse-Compton scattering. Furthermore, ambient (galactic or extragalactic) magnetic fields can induce electrons and positrons to emit in the radio band through the synchrotron process. Inverse Compton can also be responsible of an X-ray continuum emission. Therefore, WIMPs can be searched for with a full multi-wavelength and multi-messenger approach. In addition to these opportunities, that collectively represent the {\sl indirect search} technique and test the self-annihilation or decay DM processes, the local component of galactic WIMPs can also directly interact with a passive low-background detector located in underground laboratories: in this case ({\sl direct search}), the signal is the deposited energy in the detector and the tested interaction is the one between DM and ordinary matter (typically nuclei, but also electrons, especially for light DM) \cite{direct}.

Lowering the mass of the DM particle makes some channels unavailable, while others remain open. MeV DM can therefore contribute (in most models though its decay) a low-energy electron and positron flux, with its inverse Compton and synchrotron emissions producing electromagnetic radiation again from the  X ray down to the radio band. Lighter particles, like e.g. keV sterile neutrinos, can only produce photons (or standard neutrinos): the X ray and infrared bands are typically the most investigated. Even lighter particles, like axions, can only produce low-frequency photons, in the microwave or radio bands. While the CMB is a dominant radiation field at microwave frequecies, it is nevertheless well understood: a DM emission in this frequency range can potentially emerge as a distortion of the perfect CMB black-body spectrum. Astrophysical radio emission away from the CMB peak is more erratic, but DM decay contributes in a way that is morphologically and/or spectrally different from radio astrophysical sources.

CMB can be used also to study WIMPs or other heavier particles: if DM (again) annihilates or decays into electrons and positrons, these particles can distort the CMB spectrum: this can happen, for instance, in a galaxy cluster, where the large amount of DM can produce a distribution of non-thermal electrons, such that Compton scattering distorts the CMB passing through the cluster (it's a Sunyaev-Zeldovich effect due to relativistic electrons) \cite{SZ}. The same can occur if DM produces electrons in the early Universe, contributing an ionization component that can distort the CMB and be felt in the CMB temperature anisotropies at large cosmological scales \cite{CMB}. 

In most models, also neutrinos are a produced signal, but they become increasingly more difficult to detect when their energy decreases, and in some energy range they also have overwhelming backgrounds, like in the case of the 1-10 MeV range, where solar neutrinos are a dominant irreducible flux. Neutrinos as messengers of the presence of dark matter are typically (although not exclusively) investigated when they are at or above the atmospheric neutrinos energy range, and therefore when they can be produced by WIMPs or even heavier DM particles.

In the following, we will briefly report on the status of indirect search signals.

\subsection{Antinuclei}

DM annihilation (or decay) into a baryonic component can add an exotic contribution to charged cosmic rays. It is therefore convenient to look into the antinuclei production, since in the antimatter channel the astrophysical background is significantly smaller. Antiprotons are the most abundant anitinuclei component produced by DM: the spectral feature are similar to the astrophysical background (due to cosmic rays interactions in the Galaxy), and possess a high-energy cutoff that corresponds to the DM mass \cite{pbar}. The antiproton channel has been largely investigated, and currently provides interesting bounds on the DM particle physics properties. Theoretical uncertainties are present \cite{pbar}(due to imperfect knowledge on the charged cosmic rays transport in the galaxy and also to the nuclear process that leads to the formation of the antiprotons). These uncertainties limit somehow the potentiality of this search channel: nevertheless, for the most probable expectations on the cosmic-ray transport parameters, an upper bound close to 40 GeV on the DM mass for a thermal WIMP are already set by using the PAMELA data \cite{pbarbound}.

Antideuterons \cite{antiD} are produced much more rarely produced by DM, about a factor of 10$^{-4}$ less abundant than antiprotons. However, they have a very favourable signal-to-background ratio at low kinetic energies: this is due to kinematical reasons, related to the fact that DM particles annihilate almost at rest from a non-baryonic initial state, while antideuterons produced by cosmic rays come from high-energy cosmic rays impinging on the non-relativistic interstellar gas, in a process that already starts from a non-null baryonic initial state. For this reason, at kinetic energies below a few GeV per nucleon, the antideuteron flux from DM is not suppressed, while instead the background is \cite{antiD}. Even though transport in the galactic and solar environment moderates the difference between the two (DM and background) fluxes, nevertheless the handle to isolate a DM signal is relevant. Antideuterons represent a potential channel for a DM signal discovery \cite{antiD, antiD2}. In the next years, AMS-02 will provide its results on antideuteron searches in two low-energy windows. Recently GAPS has been approved by NASA, and in about 5 years will provide data in the very-favourable energy window below 300 MeV. \cite{antiD3,PR}

\subsection{Electrons and positrons}

After the PAMELA \cite{PAMELA} and AMS-02 \cite{AMS} observation, supported also by Fermi/LAT \cite{Fermi}, that above 10 GeV positrons are significantly more abundant than what is expected if they are originated purely from cosmic rays interactions in the galactic plane, a lot of discussion on their origin has arisen \cite{positrons-DM,positrons-DMnoi,positrons-astro,positrons-astronoi,positrons-other}. What is now clear is that this population of high-energy positrons requires local sources in the Galaxy. These positrons can either be supplied from DM annihilation or decay \cite{positrons-DM,positrons-DMnoi}, or from astrophysical sources like pulsar wind nebulae or other astrophysical mechanisms \cite{positrons-astro,positrons-astronoi,positrons-other}. Many thorough analyses have shown that the pulsar hypothesis can reproduce well the observed positron "excess" \cite{positrons-astro,positrons-astronoi}. DM can also reproduce this excess, although spectral features and the non-observation of an accompanying large flux of antiprotons limit the possibility to a DM particle that couples only to leptons and that it is typically very heavy \cite{positrons-DM,positrons-DMnoi}. At the same time, a such large electron and positron production from DM annihilation is seriously bounded from gamma-rays observations, since these significant leptonic component would produce gamma-rays from the inverse-Compton process on the CMB and galactic radiation fields \cite{positrons-DM,positrons-DMnoi}. A pure DM interpretation is therefore puzzling. At the same time, under the hypothesis that pulsars have a major role in the positron "excess", relevant bound on the DM properties have been set \cite{positrons-astro,positrons-astronoi}. AMS-02 already provided an incredibly accurate measure of the positron flux, allowing to reach very high-energies and showing that the positron flux relative to the total leptonic flux is actually flattening out. These are very important pieces of evidence to try to understand the origin of these positrons. The expected future releases from AMS-02 will be very important to see the behaviour of the flux also at some higher energy: this will help in understanding whether a DM emission is indeed present.

\subsection{Neutrinos}

Neutrinos can be produced by DM annihilating in the Galaxy, or in specific targets like dwarf galaxies. However, neutrinos can also be produced by DM accumulated inside the Sun or the Earth, and annihilating in their cores: this phenomenon allows us to have a close overdensity of DM, that produces a signal in a very specific place in the sky. The only particle, produced by DM, that can escape the Sun or Earth media is the neutrino: neutrino telescopes are therefore a crucial tool to investigate DM. 

For neutrinos produced inside the Sun, the relevant DM mass range falls in the WIMP sector, and requires DM to be heavier than about 5 GeV in order to be gravitationally captured (lighter particles easily evaporate on a time scale much smaller than the age of the star). In order to be captured, DM needs to be slowed down and accumulated inside the star: this occurs as a consequence of DM interactions with the Sun medium, and therefore it depends on the DM-nucleus scattering cross-section, the same probed by the direct detection technique (although in a different kinematical range). When the capture and annihilation processes equilibrate, a steady flux of high-energy neutrinos arrived to the detector from the Sun (a similar process occurs in the center of the Earth) \cite{neutrinos}. The relevant neutrino energies are the same as those of atmospheric neutrinos, which therefore represent the dominant background to this type of sources. Notice that very energetic neutrinos (above the TeV scale) are efficiently absorbed in the Sun interior, and therefore they do not contribute to the signal \cite{neutrinos}. Quite relevant bounds have been obtained recently from The Antares and IceCube Collaborations \cite{antares}. One must remember that equilibration is not necessarily achieved, especially inside the Earth: when this occurs, bounds are less stringent, even significantly.

Interesting bounds have also been obtained from dwarf galaxies, the galactic center and the Virgo cluster \cite{extrantares}. IceCube PeV neutrinos have also been discussed in terms of superheavy DM decay \cite{esmaili}, an interesting possibility that opens an observational window on a class of DM candidates that can arise for instance in grand unified theories. Good prospects are also present for future large area neutrino telescopes, like Km3NET \cite{km3} and HyperK \cite{HK} or for PINGU \cite{PINGU}.

\subsection{Gamma-rays}

Gamma rays offer a unique opportunity to study DM: being neutral, they do not suffer from the transport mechanisms that affect charged cosmic rays, and therefore they more directly trace their source morphology and spectral features. This is important for DM studies, since DM has a very specific distribution in our Galaxy and in the extragalactic environment: being able to trace direction of arrival of the photons can be used to enhance the signal-to-background ratio. Spectral information is also very important, since the expected DM fluxes are quite different from the typical power-law behaviour of astrophysical sources.

In recent years, the Fermi/LAT  has provided a remarkable amount of high-quality data, in all directions in the sky and in a wide range of energies: from a few hundred of MeV to a few hundred of GeV, a perfect range to study WIMP DM. Expected DM gamma-ray emission is morphologically quite different from the Fermi/LAT observed sky: in fact, the dominant gamma-ray emissions are the galactic foreground emission and resolved astrophysical sources. What is left after these component are subtracted (or masked) is the diffuse gamma-ray background (DGRB), which is where a DM signal is expected to hide \cite{DGRB}.  The DGRB has a spectral behaviour and an intensity which is well explained by the cumulative emission of unresolved blazars and star-forming galaxies. A typical DM energy spectrum is quite different from the DGRB measured flux: this allowed to set quite interesting bounds in the DM properties, that are approaching thermal-WIMPs cross-sections as the data collection increases \cite{DGRB}. 

A very interesting target is offered by dwarf galaxies, satellite of our own Galaxy. They are relatively close DM dominated objects: this makes them in a special position for DM studies. The analysis of the currently known dwarf galaxies is currently providing one of the best bounds on DM properties: thermal cross sections are excluded if DM is lighter than 80 GeV \cite{dwarf1}. Forecasts for the future campaigns can move this bound up by a factor of two. Nevertheless, one has to consider that uncertainties in the dwarf galaxy DM content and concentration are under debate \cite{dwarf2}.

Another very interesting target, where again DM is expected to be concentrated and therefore produce a larger gamma-ray flux, is the galactic center. While this is the place where galactic DM produces its largest signal, it is also the most complex place to look at, since it is also a very active place from the astrophysical point of view. Moreover, we live in the galactic plane, and when we look at the center we are forced to integrate everything that occurs along the galactic plane. Nevertheless, the galactic center represents a very interesting target. From one side, HESS reported very stringent bounds on DM at the TeV scale from its observations at the galactic center \cite{HESScenter}. From the other side, investigation of the galactic center at the Fermi/LAT energies has reported an "excess" on top of the expected galactic foreground \cite{GCcenter}. The "excess" has a morphology which is quite centric, and well spherical, and it appears to exceed expectations in the few to ten GeV range. This "excess" has been interpreted to be originated from one (or a combination) of three sources: a population of millisecond pulsars; emission from a series of burst-like events occurred in the past in the active galactic center environment; dark matter. The DM hypothesis is the most intriguing and it has been discussed in great details. The "excess" has, both spectrally and morphologically, the correct features expected from DM. In case it is in fact originated from DM, the DM particle needs to lie in the interesting mass range between 10 and 100 GeV range, with an almost thermal cross section  \cite{GCcenter}. On the other hand, statistical analyses the emission are more inclined toward the interpretation of a population of point-like sources: in this case, pulsars would be the obvious possibility \cite{GCstat1,GCstat2}. 

Prospects for gamma-ray investigation are bright: Fermi/LAT operations have been extended, and therefore the detector will continue to provide the high-quality data that have already instrumental to gain exceptional insight in gamma-rays astrophysics. At the same time, higher energies (in excess of 300 GeV, up to several TeV) are covered by Magic, HESS and HAWC and will be explored in great detail by CTA. These higher energies are important for DM studies, since they will cover a mass range (the TeV scale) where WIMPs are a natural solution to the DM problem (the WIMP "miracle", in its most natural implementation, requires a DM particle of either a few GeV or a few TeV), even more than the 10-100 GeV range. At GeV-TeV energies, DAMPE and in the future HERD will provide improved energy and angular resolution (very important for the statistical studies discussed below, provided that the photon statistics is large enough). Lower energies, in the MeV to GeV range, could be covered by AstroGam and PANGU: this energy range is interesting for the subGeV to few GeV DM, and to explore the low tail of multi-GeV WIMP DM, useful to pinpoint the spectral feature of the DM emission, a feature that correlated to the way DM annihilate (into which final states) and therefore to how it is coupled to ordinary matter.

\subsection{Radio and microwave}

Radio frequencies have been shown to be another interesting opportunity for DM searches, including WIMPs \cite{radio1,radio2}. For electrons and positrons of GeV-TeV energies (as those produced by WIMP annihilation) in a microGauss magnetic field (the typical order of magnitude in galaxies and clusters), synchrotron emission falls in the MHz to GHz range. This is the range where radioastronomy operates: available data have already been used to set bounds on DM \cite{radio1}, with the interesting result that these bounds are similar to those obtained by the other indirect search techniques. Bounds from galactic DM emission from all sky-maps limit a DM WIMP to be heavier than 10 GeV, and in deriving this bound the interplay of the different frequency survey is instrumental (lower frequencies are better suited to constrain lighter DM). Extragalactic radio emission has a similar constraining power under reasonable assumptions, but it is affected by larger uncertainties arising from DM modelling and the actual size of ambient magnetic fields, which is affected by large uncertainties, while being a crucial ingredient in the theoretical prediction for the DM radio signal. Specific targets, like dwarf galaxies or the Andromeda galaxy \cite{target}, provide a good opportunity, since they are relatively close, and, due to the exceptional angular resolution of radio surveys, they can be deeply probed.

An interesting feature of radio observation is that apparently an excess of radio emission is observed. This result, often called the "Arcade excess" \cite{arcade1} has been tested and confirmed. The origin is currently not understood, and it has been shown that is has the correct features of a DM signal produced by an almost-thermal WIMP with a mass in the 10-100 GeV range \cite{arcade2}. The result is intriguing, although it can be challenged by anisotropy measurements \cite{arcade3}.

The CMB, as already discussed above, offers an interesting tool to set bounds on DM \cite{CMB}.
Here I will only mention the very solid (and stringent) bounds that can be derived from the CMB as a consequence of injection of ionising
particles during the cosmic dark ages: DM annihilation increases the residual ionisation of the Universe, and this affects the CMB temperature spectrum and anisotropies. Thermal DM with masses below 10-20 GeV are excluded \cite{CMB}.

\subsection{Low-frequencies and non-WIMP}

Non-WIMP DM can cover an immense range of masses: if these particles decay (or annihilate) in a non-relativistic regime, they produce photons with energies smaller than their own mass. In case of direct decay into a photon pair, the emission is a line (this can occur also for WIMPs, but it's a very suppressed process, occurring only at the loop level). Considering that axions can be as light as the $\mu$eV scale (or even lower, in principle) and that non-thermal warm dark matter can be as "heavy" as the keV scale, this implies that we can look in a very wide range of the electromagnetic spectrum. 

Here I will just report two examples. One, that has raised interest, refers to the X ray band. Observations on 73 galaxy clusters, as well as observartions of the Perseus cluster and Andromeda galaxy, reported the presence of an unaccounted for line a 3.5 keV energy \cite{xray}. If interpreted
as a DM signal, this would point to a DM particle with a mass of 7 keV. This could correspond, e.g., to a sterile neutrino DM, decaying into 2 photons. This interpretation requires that the sterile neutrino responsible for the DM is very weakly mixed with ordinary neutrinos, with a mixing angle of $10^{-5}$. The DM interpretation has been opposed, and recently new bounds on the search for sterile neutrinos have been reported from nuSTAR that makes the DM interpretation marginal \cite{nustar}.

As another example on how low frequencies can be used to study DM, and more generally new light particles with an astroparticle physics approach, let me mention the excess in the near infrared background that can be a hint of a eV-mass DM emission \cite{nirb}, or the bounds on particles with 10 meV mass from the cosmic infrared background measure by Spitzer and FIRAS \cite{cib}. Light particles have gained interest recently as an alternative to the WIMP paradigm: astrophysics can provide remarkable insight in constraining these hypothesis (or to identify a signal).

\subsection{Going beyond: statistical correlations}

Being the sum of many independent sources (either DM halos or astrophysical sources), the DGRB is to first approximation isotropic. But at a deeper level anisotropies are clearly present. This means that, even though sources are too dim to be individually resolved, they can affect the statistics of photons across the sky. Recently, different techniques have been proposed to investigate the unresolved components of the gamma-rays sky. DM emission is actually among the dimmest contributors: these new methods, based on statistical analyses, can therefore be a turning point in the identification of a signal.

The first method is based on a 1-point statistics estimator, the so-called "photon pixel count", based on the fact that different sources, with varying levels of emission and distribution in the sky, contribute a different number of photons to the various pixels of a gamma-rays map \cite{1pdf}. This technique has been used on the Fermi/LAT data, to derive the number count distribution of faint gamma-ray point-sources: this is a first step that allows to better understand the astrophysical populations of gamma-rays emitters, and will now be extended to DM investigation. The same technique has also been used to test whether the galactic center gamma-rays "excess" is originated by a a true diffuse emission (which would point toward a DM interpretation) or instead if it is compatible with a distribution of point sources \cite{GCstat1}, and it has been proposed for high-energy neutrinos \cite{feyereisen}.

The second technique looks at the Fermi/LAT gamma-ray sky map by constructing its 2-point correlator \cite{auto1,auto2}. The gamma-ray angular power spectrum has been measured, including its energy dependence. It appears compatible with a population of point sources just below the detection threshold. The energy dependence also points to a hint for the presence of two different source populations, one dominating below 10 GeV and one above. Being compatible with point sources, no evidence of a DM signal is present in the angular power spectrum: this allowed to set bounds on WIMP DM \cite{auto2}. These bounds less constraining than the bounds coming directly from the integrated DGRB, but are obtained from an observable that feels how DM is distributed in the Universe.

A potentially more powerful technique is the cross-correlation between an electromagnetic DM signal (e.g., gamma rays) and a gravitational tracer of DM in the Universe \cite{cross-shear,cross-lss,cross-cmb}. This technique has been proposed recently and it exploits two distinct features of particle DM, one referring to its particle physics nature and one that is a direct evidence of the presence of DM. The best gravitational tracer is a lensing observable, since it measures directly, in an unbiased way, the presence of DM and where it is. Cosmic shear is the most direct option, since it directly traces the whole DM distribution. CMB lensing traces DM imprints on the CMB anisotropies, and is more sensitive to high redshift. Alternatively,  gravitational tracers can be galaxy or cluster catalogs. In this case, the gravitational tracer is biased, since it traces light and not directly DM, but on the other hand the statistics of detected objects can be large, and it will be even more so in the future, since galaxy surveys are one of the major current and planned endeavours in cosmology. Auto- and cross-correlations have been recently discussed also in the X-ray band, in connection to the search for DM composed by sterile neutrinos at the keV mass scale \cite{zandanel}.

A positive cross-correlation between Fermi/LAT gamma-ray maps and a host of galaxy catalogs (2MASS, SDSS, NVSS) has been detected \cite{cross-meas-lss1}. This can be very well explained by DM, with a mass (around 100 GeV mass) and interactions (close to thermal) in the interesting range form WIMPs \cite{cross-meas-lss2}
Although at the moment it's not statistically possible to disentangle a pure DM contribution from a blazars component, nevertheless, the results show that this technique is potentially able to identify a DM signal even though the total gamma-ray DM emission is largely subdominant in the total DGRB emission. This shows the potentiality of the cross-correlation technique \cite{cross-meas-lss2}. More recent investigations can be found in Ref. \cite{cross-meas-lss3}.

Recently, a positive signal has been measured also by correlating cluster catalogs with the Fermi-/LAT gamma rays maps \cite{cross-meas-cluster}. Concerning the cross-correlation with weak lensing observables, at the moment the cosmic shear catalogs are still covering small portions of the sky and no signal has been detected \cite{cross-meas-shear}. In the next years, especially with the Dark Energy Survey (DES) data and, in a farther future, with Euclid, this technique will definitely be able to detect a positive correlation, which will then be posed under scrutiny to identify whether it refers to astrophysical sources (therefore gaining insight in the type and distribution of gamma-rays emitters) or if instead a DM component is present. It has been shown that under favourable conditions, DES could be able not only to identify a signal, but also to infer the particle physics parameters of DM (its mass and interactions) \cite{cross-shear}. Euclid will then have the possibility to either pinpoint the DM particle properties or to significantly disprove the WIMP hypothesis  \cite{cross-shear}. What is clear is that we have to exploit all opportunities: one option does not fit all possibilities, and to understand what DM is requires to plan with the idea to explore all alternatives.
A positive cross-correlation has also been detected between Gamma-rays and CMB lensing: this correlation arises from higher redshift, as compared to the one measured with galaxy catalogs. Since DM emits its electromagnetic signals mostly at low redshift, galaxy and cluster catalogs and cosmic shear are the places where to put under scrutiny the DM hypothesis; CMB lensing instead can provide a leverage to constrain the astrophysical emission, and therefore reduce the uncertainty on this component in the large-scale-structure and weak-lensing cross-correlation.

\section{Conclusions}

Multimessenger astrophysics offers a wide range of opportunities to study DM in most its full mass range: from axions and axion-like particles, to sterile neutrinos and majorons, to WIMPs and non-thermal alternatives in the GeV to TeV mass range, all the way to super-heavy DM. ˜Astrophysical fore/back-grounds are very complex and typically dominant over the sought-after DM signals: this implies that
˜a clever identification of potential targets, signatures and signal features is necessary. DM searches will in fact progress together with a better understanding and modelling of the astrophysical environment, and together with the identification and development of new
methods and techniques (like, e.g., the statistical approach of cross-correlations, recently proposed). Up to know, some 
intriguing hints have been identified, but to establish that any of these clues are indeed originated by DM will require further
investigation, deeper understanding of the astrophysical settings and, possibly, independent confirmation in a different
exploration channel. The field is progressing rapidly and will profit from a large wealth of data expected in the next 5-15 years, that
will make the investigation of DM as an elementary particle even more exciting and hopefully conclusive. 
\bigskip 
\begin{acknowledgments}
Work supported by the research grant {\sl Theoretical Astroparticle Physics} number 2012CPPYP7 under the program PRIN 2012 funded by the Ministero dell'Istruzione, Universit\`a e della Ricerca (MIUR), by the research grants {\sl TAsP (Theoretical Astroparticle Physics)} and {\sl Fermi} funded by the Istituto Nazionale di Fisica Nucleare (INFN).

This contribution is dedicated to Prof. Vera Rubin, who sadly passed away on December 25, 2016. Vera Rubin has been a true pioneer
in the study of dark matter, and a sincere source of inspiration to all of us working in this field.

\end{acknowledgments}

\bigskip 

\end{document}